\documentclass[amsmath,nofootinbib,twocolumn,superscriptaddress]{revtex4-1}
\pdfoutput=1
\usepackage[colorlinks=true,citecolor=blue,urlcolor=blue]{hyperref}
\usepackage{aas_macros,graphicx,marvosym,subfigure,amssymb,amsmath}
\usepackage{comment}
\usepackage{slashed}

\newcommand{\pd}{\,\partial}

\newcommand{\SG}[1]{\textcolor{red}{\textsf{[SG: #1]}}}
\newcommand{\NI}[1]{\textcolor{blue}{\textsf{[NI: #1]}}}

\newcommand{\be}{\begin{equation}}
\newcommand{\ee}{\end{equation}}
\newcommand{\bea}{\begin{eqnarray}}
\newcommand{\eea}{\end{eqnarray}}

\newcommand{\bln}{\begin{align}}
\newcommand{\eln}{\end{align}}
\newcommand{\bst}{\begin{split}}
\newcommand{\est}{\end{split}}
\newcommand{\bi}{\begin{itemize}}
\newcommand{\ei}{\end{itemize}}
\newcommand{\ben}{\begin{enumerate}}
\newcommand{\een}{\end{enumerate}}

\def\le{\left}
\def\ri{\right}
\def\ha{{1\over 2}}

\def\al{{\alpha}}

\def\det{{\rm det}}

\def\th{{\theta}}

\def\ep{{\epsilon}}

\newcommand{\p}{\partial}

\newcommand\ga{{\ensuremath{{\gamma}}}}

\newcommand\sig{\sigma}

\newcommand\lam{\lambda}
\newcommand\Lam{\Lambda}

\newcommand\Om{\Omega}

\def\lam{{\lambda}}

\def\eeq{\end{equation}}

\newcommand\sL{{\ensuremath{{\mathcal L}}}}

\newcommand\bpsi{{\bar \psi}}

\def\th{{\theta}}

\newcommand\vep{{\varepsilon}}

\begin{document}

\title{Effective Field Theory of Force-Free Electrodynamics}

\author{Samuel E. Gralla}
\email{sgralla@email.arizona.edu}
\affiliation{Department of Physics, University of Arizona, Tucson, Arizona 85721, USA}

\author{Nabil Iqbal}
\email{nabil.iqbal@durham.ac.uk}
\affiliation{Centre for Particle Theory, Department of Mathematical Sciences, Durham University,
South Road, Durham DH1 3LE, UK}

\begin{abstract}
Force-free electrodynamics (FFE) is a closed set of equations for the electromagnetic field of a magnetically dominated plasma.  There are strong arguments for the existence of force-free plasmas near pulsars and active black holes, but FFE alone cannot account for the observational signatures, such as coherent radio emission and relativistic jets and winds. We reformulate FFE as the effective field theory of a cold string fluid and initiate a systematic study of corrections in a derivative expansion.  At leading order the effective theory is equivalent to (generalized) FFE, with the strings comprised by magnetic field line worldsheets.  Higher-order corrections generically give rise to non-zero accelerating electric fields ($\mathbf{E}\cdot \mathbf{B}\neq 0$). We discuss potential observable consequences and comment on an intriguing numerical coincidence.
\end{abstract}

\maketitle

\section{Introduction}

The astronomical universe abounds with spectacular phenomena that defy explanation years or even decades after discovery.  
Among these persistent puzzles are the origin of ultra high-energy comsic rays, the mechanism of coherent radio emission (from pulsars and fast radio bursts), and the formation of relativistic particle jets and winds \cite{kotera-olinto-review2011,magnetoluminescence,katz-review2016,beskin-review2018}.  Simple energetics implicates compact objects (neutron stars and black holes) in all of these phenomena, but a complete theoretical treatment has remained elusive.  The foundations are secure \cite{goldreich-julian1969,blandford-znajek1977}: rapid rotation and strong magnetic fields give rise to diffuse plasma that efficiently carries energy away from the central object.  But how does this energy get converted into the signals we see?

The energy-carrying plasma is elegantly described by the theory of force-free electrodynamics (FFE) \cite{uchida1997general,komissarov2002,Gralla:2014yja}.  The assumption is that charged particles are sufficiently sufficiently plentiful that they screen the electric field (setting $\mathbf{E}\cdot\mathbf{B}=0$ with $B^2>E^2$), but sufficiently diffuse that they exchange little stress-energy with the fields (setting $\mathbf{j}_{\rm el} \cdot \mathbf{E}=0$ and $\rho_{\rm el} \mathbf{E}+ \mathbf{j}_{\rm el} \times \mathbf{B} =0$).  These conditions are expressed covariantly as
\begin{subequations}\label{constraints}
\begin{align}
\epsilon^{\mu \nu \rho \sigma}F_{\mu \nu} F_{\rho \sigma} & = 0 \qquad \textrm{(degenerate)} \label{degenerate} \\
\quad F^{\mu \nu} F_{\mu \nu} & >0 \qquad \textrm{(magnetically dominated)} \label{magdom}
\end{align}
\end{subequations}
as well as
\begin{align}
j^{\mu}_{\mathrm el} F_{\mu\nu} & = 0 \qquad \textrm{(force-free)}. \label{force-free}
\end{align}
When combined with Maxwell's equations ($j^{\mu}_{\mathrm{el}} = \nabla_{\nu} F^{\mu\nu}$ and $\nabla_{[\mu} F_{\rho\sig]} = 0$), the force-free condition \eqref{force-free} becomes 
\begin{align}
\nabla_{[\mu} F_{\rho\sig]} = 0, \qquad F_{\sig\nu} \nabla_{\mu} F^{\mu\nu} = 0.\label{FFE}
\end{align}
Remarkably, Eqs. \eqref{FFE} and \eqref{constraints} comprise a  well-posed (hyperbolic) evolution system \cite{komissarov2002,palenzuela-etal2011,pfeiffer-macfadyen2013,carrasco-reula2016}: these non-linear equations can be used to evolve the electromagnetic field forward in time in a self-consistent manner, while making no reference to the dynamics of the charges themselves. 

There are strong theoretical arguments that active pulsars and black holes possess force-free plasmas (see \cite{goldreich-julian1969,blandford-znajek1977} and many later references).  Famously, this enables efficient extraction of the rotational energy of the compact object.  However, a purely force-free magnetosphere cannot accelerate particles into jets, winds, or cosmic rays (it has $\mathbf{E}\cdot \mathbf{B}=0$ exactly), and it cannot radiate at any wavelength not put in by initial/boundary conditions (as a theory with no intrinsic scale).  In other words, FFE alone is too simple to account for observations.

The most common approach to going beyond FFE is to invoke large departures in small regions, such as particle acceleration in a reconnecting current sheet.  While this kind of violation is undoubtedly part of the story, the many remaining puzzles---notably the lack of a coherent radiation mechanism---have motivated us to try the opposite tack of studying \textit{small} departures in \textit{large} regions.  Plausibly, a numerically small correction could become important via the growth of an instability---perhaps of exactly the clumping character needed to provide the observed radio emission.  Similarly, even a small amount of $\mathbf{E}\cdot \mathbf{B}$ could assist with explaining the origin of pulsar winds 
or the loading of relativistic jets in active galactic nuclei, while an anomalously large correction could account for high-energy cosmic rays.


We therefore propose---and begin---a systematic study of \textit{corrections to force-free electrodynamics}.  While ultimately such corrections ought to be derivable from the fundamental theory of quantum electrodynamics (QED), in practice this is difficult even for the leading force-free behavior.  We will therefore resort to the framework of effective field theory (EFT), based on the mantra that whatever \textit{can} appear, \textit{will} appear.  The effective field theorist need only identify the symmetries that characterize the relevant phase (in this case a cold, strongly magnetized plasma), after which she simply writes down every allowed correction.  Each term will come with a coefficient of undetermined size that could in principle be computed from microscopics.  If the symmetries are realized correctly, then by definition the needed correction is \textit{somewhere} in this list, and an exhaustive study of the phenomenology is bound to find it eventually.


In fact, the first steps in this program of correcting FFE have already been taken in disguise.  In Ref \cite{Grozdanov:2016tdf}, a reformulation of magnetohydrodynamics (MHD) was presented where it was argued that the theory could be efficiently reorganized from a strictly symmetry-based point of view, taking as fundamental starting principles not Maxwell's equations but rather the conservation of stress-energy and magnetic flux,
\be\label{currents}
\nabla_{\mu} T^{\mu\nu} = 0 \qquad \nabla_{\mu} J^{\mu\nu} = 0,
\ee
where $T_{\mu \nu}$ is symmetric and $J^{\mu \nu}$ is the dual of the usual field strength
\be\label{Fdef}
J^{\mu\nu} = \ha \ep^{\mu\nu\rho\sig} F_{\rho\sig}.
\ee
The authors noted that a consistent zero-temperature realization is possible; this turns out to be a generalized form of FFE studied previously in \cite{Freytsis:2015qda}.  It can also be seen directly that Eqs.~\eqref{currents} correspond to FFE when $T^{\mu \nu}$ is the Maxwell stress-energy tensor and $J^{\mu \nu}$ is the dual field strength.\footnote{Conservation of $J$ is equivalent to $dF=0$ or $F=dA$.  It is easy to check that $\nabla_\mu T^{\mu \nu}=-F_{\mu \nu} \nabla_\rho F^{\rho \nu}$ then holds as an identity.}

The program of correcting FFE can thus be organized as obtaining sets of closed, consistent equations respecting the conservation laws \eqref{currents} and reducing to FFE in some limit. In principle the approach of Ref.~\cite{Grozdanov:2016tdf} already allows one to study corrections to FFE; however the methods are somewhat cumbersome, as their formalism required certain constraints to be satisfied off-shell, and it proved difficult to maintain these constraints at higher orders in the derivative expansion. The main technical advance in this paper is the identification of an action principle whose field equations are precisely the conservation laws \eqref{currents}.  Using a generalization of ideas used in hydrodynamics \cite{Dubovsky:2011sj}, we promote the field line worldsheets of FFE to true dynamical degrees of freedom and assign to each a  ``worldsheet magnetic photon'' that accounts for the conserved flux. We can thereby invent consistent theories by writing down scalars.

Using this approach, we find that the unique scale-free theory at leading (``ideal'') order in derivatives is precisely FFE.  The most general ideal theory is that studied previously in \cite{Freytsis:2015qda} as generalized FFE and in \cite{Grozdanov:2016tdf} as zero-temperature magnetohydrodynamics.  At higher order in derivatives there are a variety of corrections, and we focus on those that give rise to non-zero $\mathbf{E} \cdot \mathbf{B}$.  Interestingly, these corrections are ``topological'' in that they affect only the field strength without affecting the equations of motion.  This nevertheless has observable consequences, since it is the field strength that accelerates particles.  We consider one simple such correction in detail and estimate the size required to account for pulsar winds.  Tantalizingly, the needed lengthscale is similar to the wavelength of coherent radio emission, suggesting that one lengthscale could perhaps account for both.



In addition to practical astrophysical consequences, we also hope that the symmetry-based approach to FFE will shed more light on its regime(s) of validity.  Force-free fields arise in a wide variety of physical circumstances: Besides the compact object magnetospheres considered here, they also occur in the solar corona \cite{wiegelmann-sakurai-review2012} and even in relaxed laboratory plasmas \cite{gray-brown-dandurand2013}.  These three types of plasma are in entirely different physical regimes, and indeed entirely different physical arguments converge on the force-free description in each case.  (We have presented only the argument relevant to compact objects.)  A helpful analogy here is perhaps the Landau theory of Fermi liquids (see e.g. \cite{nozieres1999theory}): a wide class of metals are described by Fermi liquids not because interactions are always weak, but rather because Fermi liquid theory can be formulated as an effective theory with (almost) no relevant operators \cite{Polchinski:1992ed,RevModPhys.66.129}. Here we suggest that a similar argument provides a basis for the ubiquity of FFE.

In Sec.~\ref{sec:symmetries} we review the philosophy of effective field theory, identify the relevant microscopic symmetries of QED, and motivate the emergent symmetries of our effective action.  Readers uninterested in motivation may skip to Sec.~\ref{sec:action}, where prove that the field equations are conservation laws and study the derivative expansion.  In Sec.~\ref{sec:solutions} we consider some simple solutions.  Finally in Sec.~\ref{sec:observations} we attempt to connect with observations.

Our metric $g_{\mu \nu}$ has signature $(-+++)$.  The spacetime volume element is denoted $\epsilon^{\mu \nu \rho \sigma}$ ($\star$ is the Hodge dual), with $\varepsilon^{\mu \nu}$ referring to an induced element on a submanifold.  We use Heaviside-Lorentz units with $\hbar=c=1$.

\section{Symmetries}\label{sec:symmetries}

\subsection{Microscopic symmetries of the system} 

The microscopic description of our system is  presumably QED, which we regard as non-perturbatively defined by the path integral.  A full solution of QED would entail a complete knowledge of this path integral as a function of external sources, from which any desired correlation function can be computed by differentiation. Here we are interested in sources associated with conserved quantities, as only these are expected to have a simple universal description. 

As mentioned in the introduction, there are two conserved currents of interest.  One is the stress tensor $T^{\mu\nu}$, whose source is the metric, and whose conservation follows as usual from diffeomorphism invariance. 

The other is the two-form current $J^{\mu\nu}$ that measures magnetic flux. Interestingly, the symmetry principle behind the conservation of such a higher-form currents has only recently been studied systematically, and is called a {\it generalized global symmetry} \cite{Gaiotto:2014kfa}. Such symmetries have  found applications in diverse physical contexts, ranging from constraining the phase structure of gauge theories and topological phases \cite{Gaiotto:2017yup,Komargodski:2017dmc,PhysRevB.93.155131,Wang:2014pma,Tanizaki:2017mtm,Kitano:2017jng,Gaiotto:2017tne} to an understanding of gauge bosons as Goldstone modes \cite{Gaiotto:2014kfa,Hofman:2018lfz,Lake:2018dqm} to symmetry-based formulations of hydrodynamics  \cite{Grozdanov:2016tdf,Grozdanov:2017kyl,Grozdanov:2018fic,Armas:2018zbe,Armas:2018atq,Grozdanov:2018ewh,Glorioso:2018kcp}.  The realization of the generalized global symmetry associated with the conservation of magnetic flux will be a technical tool in our analysis (we call it the ``magnetic photon shift''); for now we simply note that the source for the 2-form current $J^{\mu \nu}$ is a fixed 2-form classical field that we call $b_{\mu \nu}$. 
 
We therefore take the partition function to depend on these sources:
\begin{align}
Z[g,b] = \int [d\psi dA] \exp\le(i S_{QED}[\psi, A; g, b]\ri),\label{ZQED}
\end{align}
where $S_{QED}$ is the microscopic QED action: 
\begin{align}
S_{\tiny \rm QED} & = \int d^4x \sqrt{-g} \bigg(-\frac{1}{4}(dA)^2+ \bpsi \le(\slashed{D} + m \ri) \psi \nonumber \\ & \qquad \qquad \qquad \qquad +  \frac{1}{4} b_{\mu\nu} \ep^{\mu\nu\rho\sig} (dA)_{\rho\sig}\bigg). \label{qedac} 
\end{align}
The associated correlation functions are (by definition) the stress-energy and two-form current,
\be
\langle T^{\mu\nu} \rangle \equiv \frac{2}{\sqrt{-g}} \frac{\delta W[g,b]}{\delta g_{\mu\nu}} \qquad \langle J^{\mu\nu} \rangle \equiv \frac{2}{\sqrt{-g}} \frac{\delta W[g,b]}{\delta b_{\mu\nu}}. \label{TJdef}
\ee
where $W = -i \log Z$.  In these equations it is implied that, after variation, one sets $g_{\mu \nu}$ and $b_{\mu \nu}$ to values corresponding to the background of interest.  The metric of course describes external gravitational fields, while $b$ describes the external charge density $j_{\rm ext}=\star db$,
\begin{align}
    j^{\sig}_{\mathrm{ext}} = -\frac{1}{2}\ep^{\sig\rho\mu\nu} \p_{\rho} b_{\mu\nu}. \label{jdef} 
\end{align}

We then \textit{define} the field strength $\langle F_{\mu \nu} \rangle$ by $\star F=J$,
\begin{align}
    \tfrac{1}{2} \epsilon^{\mu \nu \rho \sigma} \langle F_{\rho \sigma} \rangle = \langle J^{\mu \nu} \rangle.
\end{align}
This agrees with the usual definition $F=dA$ when the dynamical variable is the gauge field $A$, but will more generally apply when other dynamical fields are used.  The field strength behaves as usual with respect to external charges, pushing them around by the Lorentz force law (App.~\ref{app:externalcharge}).

For our purposes a symmetry of the theory is captured\footnote{The generalized global symmetry in question is represented on the dynamical fields by a transformation parametrized by a closed but not exact $1$-form $\Lam$. However, this can be promoted to an invariance of the partition function under a transformation by an {\it arbitrary} $1$-form $\Lam$ by shifting $b$ by $d\Lam$ to compensate, as in \eqref{inv-b}; see Appendix \ref{app:magphot} for more discussion of this point.} by an invariance of the partition function with respect to a transformation of the sources.  From \eqref{ZQED}, we see that QED has the symmetries
\begin{subequations}\label{inv}
\begin{align}
Z[\phi^* g, \phi^* b] & = Z[g,b] \label{inv-g} \\
Z[g, b+d\Lam] & = Z[g,b], \label{inv-b} 
\end{align}
\end{subequations}
where $\phi^*$ is the action of a diffeomorphism and $\Lambda$ is an arbitrary 1-form. Note that the $U(1)$ gauge symmetry  of electromagnetism as written in \eqref{qedac} does not result in such an invariance of the partition function. 

To each symmetry is associated a conservation law.  Varying with respect to an infinitesimal 1-form shift $b \to b+d\Lambda$, from \eqref{inv-b} we find $\nabla_{\mu} \langle J^{\mu \nu} \rangle =0$. Varying with respect to an infinitesimal diffeomorphism and using this result, it follows from \eqref{inv-g} that
\be
\nabla_{\mu} \langle T^{\mu\nu} \rangle = \ha (db)^{\nu}_{\phantom{\nu}\rho\sig}\langle J ^{\rho\sig} \rangle. \label{consT}
\ee
The right-hand-side reflects non-conservation in the presence of an external electric current $db$.  Thus in the absence of external electric charge, the symmetries \eqref{inv} imply the conservation of the correlators \eqref{TJdef}, 
\be\nabla_{\mu} \langle T^{\mu\nu} \rangle = 0, \qquad \nabla_{\mu} \langle J^{\mu\nu} \rangle = 0. \label{consTJ} 
\ee



This presentation of QED minimizes the importance of the particular degrees of freedom $(\psi, A_\mu)$ that are integrated over in the path integral. While these degrees of freedom are weakly coupled near the vacuum of empty space, they are not necessarily the ideal degrees of freedom for describing a strong-field plasma.
The idea of EFT is that there 
should exist a new set of fields $\Phi$ with a new local action $S_{\rm EFT}$ that reproduces the partition function in the regime of interest,
\begin{align}
Z[g,b] \approx \int [d\Phi] \exp\le(i S_{\rm EFT}[\Phi; g, b]\ri),\label{ZZ}
\end{align}
while also providing a simpler description in this regime.
We have chosen to regard \eqref{ZQED} as exact and \eqref{ZZ} as approximate, but most properly we only require an overlapping regime of validity.\footnote{Opinions may differ as to which theory is more fundamental.  The civilization on the crab pulsar likely studies quantum electrodynamical phenomena in terms of some action with degrees of freedom relevant to strong-field plasma, and they wonder about alternative actions. 
Without guidance from expensive particle accelerators that artificially construct regions of weak magnetic fields, they have little hope of constructing the full theory we call QED, but they do make effective actions as they ponder what life might be like on the surface of the distant planet Earth!} 
As a practical matter, our task is to identify appropriate degrees of freedom $\Phi$ for a strong-field plasma, subject only to the restriction that the partition function respects the symmetries \eqref{inv}.

\subsection{Emergent symmetries of description}

Our task now is to identify degrees of freedom relevant to the macroscopic description of strong-field plasma and express them as a list of fields subject to certain local symmetries. Presumably different choices result in different phases that we can attempt to link to precise microscopic models. We have made a choice which reproduces (at leading order) the expected phenomenology of strong-field plasma; as we discuss, this is a natural generalization of the effective action for hydrodynamics of \cite{Dubovsky:2011sj} to the case of higher-form symmetry, with a slight enlargement of symmetries that we discuss below. 

We begin, however, by providing an independent physical motivation. 
Note first that while the full theory has a conserved field line \textit{number} (the magnetic flux), it offers no provision for tracking \textit{individual} field lines in time.  There is simply no way, in general, to say which field line at some later time is the  ``same'' one as at some earlier time.  However, in some regimes of plasma physics we can attempt this identification by following the motion of individual charges attached to the lines.\footnote{Classically, a particle in a strong magnetic field executes gyrations about a moving ``guiding center'', whose position in time can be used to say which field line is the ``same'' one it started on.  In reality, synchrotron radiation will quickly relax particles into the lowest Landau level.  Presumably there is an analogous story: the particle remains adiabatically in this state as the field evolves on macroscopic time and lengthscales, and its wavefunction can be used to identify the field line.}  Our EFT will integrate out the charges entirely, but we will retain this vestige of their existence by nevertheless taking the degrees of freedom to be strings. At leading order these strings will be the field lines of a flux-conserving magnetic field, but at higher orders they will not precisely align with the physical flux $J^{\mu\nu}$.

We label each string by a pair of numbers $\Phi_1$ and $\Phi_2$ (this could be the $x,y$ position where the string pierces a fiducial surface at a fiducial time), which are promoted to spacetime fields $\Phi^{I}(x)$ whose simultaneous level sets are the string  worldsheets.  This defines a \textit{foliation} of spacetime into two-dimensional string worldsheets, which we take to be regular and timelike.  
We do not want any {\it preferred} strings in our theory, so the labels $\Phi_1$ and $\Phi_2$ should be arbitrary.  We therefore require invariance under smooth relabelings\footnote{This symmetry is larger than those used in the effective action approach to hydrodynamics \cite{Dubovsky:2011sj}, which  restricted instead to volume-preserving diffeomorphisms. } (diffeomorphisms on the manifold of $\Phi_1$ and $\Phi_2$),
\begin{align}\label{relabel}
\Phi_1 \to \Phi_1'(\Phi_1,\Phi_2), \quad \Phi_2 \to \Phi_2'(\Phi_1,\Phi_2) 
\end{align}
such that
\begin{align}\label{jac}
    \det \frac{\pd(\Phi_1,\Phi_2)}{\pd(\Phi_1',\Phi_2')} \neq 0.
\end{align}
The idea of taking a foliation as a fundamental degree of freedom was discussed before in \cite{compere-gralla-lupsasca2016}.

We will now need another degree of freedom to keep track of the conserved flux mandated by \eqref{consTJ}. 
One option is to relax the full relabeling invariance \eqref{relabel} to volume-preserving diffeos; this makes $\Phi_1$ and $\Phi_2$ ``Euler potentials'' \cite{carter1979,uchida1997general,Gralla:2014yja} for the magnetic field, assigning to each string a conserved flux proportional to $|d\Phi_1 \wedge d\Phi_2|$.  
However, this is too restrictive for our purposes; we merely want a conserved flux, independent of any assignment to the strings.  

We will instead track the flux by introducing a 1-form field $a_\mu$, which (for reasons to be explained) we will call the {\it worldsheet magnetic photon}. 
As $a$ is related to the conservation of magnetic flux, it should transform under the 1-form general globalized symmetry parametrized by $\Lam$ in \eqref{inv-b}. The simplest such transformation is: 
\be
a \to a + \Lam \qquad b \to b + d\Lam, \label{ba}
\ee
where we have also recalled the transformation of the external source $b$. Note this means that the combination
\be
da - b
\ee
is invariant.\footnote{Since $a_\mu$ is shifted by a field-independent 1-form $\Lambda_\mu$, an action $S_{\rm new}$ respecting \eqref{ba} will produce a partition function \eqref{ZZ} with the desired symmetry \eqref{inv-b}.} Without a field transforming non-linearly in this manner, we would not be able to couple $b$ to any light degrees of freedom in the action. We call this the {\it magnetic photon shift}, and in Appendix \ref{app:magphot} we review why this transformation deserves this name. We emphasize that this field is not the original electric photon $A$.


To confine the magnetic photon to the worldsheet we further demand invariance under shifts by a (possibly different) 1-form on each sheet,
\begin{align}\label{chem-shift}
    a_\mu & \to a_\mu + \omega_\mu(\Phi_1,\Phi_2).
\end{align}
This is a 1-form generalization of the ``chemical shift'' of \cite{Dubovsky:2011sj}; note that as $\Phi_{1,2}$ label the  worldsheets, demanding invariance under this symmetry implies that only the variation of $a$ within each sheet can affect the dynamics.  It implies that invariant local quantities must be constructed from the worldsheet field strength [Eq.~\eqref{thingy} below] and its worldsheet derivatives, as elaborated on in Sec. \ref{sec:invob} below.


\section{Action}\label{sec:action} 

To summarize Sec.~\ref{sec:symmetries} above, we consider a theory of two scalars $\Phi_1, \Phi_2$ and a vector $a_{\mu}$ in the presence of fixed sources $g_{\mu \nu}$ and $b_{\mu \nu}$.  We assume invariance under spacetime diffeomorphisms as well as 
\begin{enumerate}
    \item String relabeling: $\Phi_I \to \Phi_I'(\Phi_1,\Phi_2)$ 
    \item String-dependent shifts: $a \to a + \omega(\Phi_1,\Phi_2)$ 
    \item Magnetic photon shift: $ a \to a + \Lambda, \ b \to b+d\Lambda$ 
\end{enumerate}

For formulating the theory we will assume that the foliation is regular and timelike, which may be expressed as
\begin{itemize}
    \item $d \Phi_1 \wedge d\Phi_2$ must be non-zero and spacelike.
\end{itemize}
We shall see that, physically, this assumption corresponds to magnetic domination \eqref{magdom}.  Its potential violation is connected with the breakdown of the theory near current sheets \cite{spitkovsky2006}, magnetic null points \cite{lyutikov-sironi-komissarov-porth2017}, or in the presence of turbulence \cite{zrake-east2016}; these interesting phenomena are beyond the scope of the present study.


\subsection{Invariant objects} \label{sec:invob}
We now construct objects invariant under the symmetries. We first introduce a more invariant description of the foliation.  The binormal field $n_{\mu \nu}$ is given\footnote{In components we have $n_{\mu\nu} = \frac{S_{\mu\nu}}{s}$ where $S_{\mu\nu} \equiv \nabla_{[\mu} \Phi_1 \nabla_{\nu]} \Phi_2$ and $s \equiv \sqrt{\frac{S^{\mu\nu}S_{\mu\nu}}{2}}$. }
by
\begin{align}
    n = \frac{d \Phi_1 \wedge d \Phi_2}{|d\Phi_1 \wedge \Phi_2|}.
\end{align}
The dual of the binormal is the induced volume element $\varepsilon$ on the foliation 
\begin{align}
    \varepsilon_{\mu \nu} = \tfrac{1}{2} \epsilon_{\mu \nu \rho \sigma} n^{\rho \sigma}. \qquad (\varepsilon=\star n) \label{epdef}
\end{align}
These forms satisfy
\begin{align}
    n_{\mu \nu} n^{\mu \nu} = 2, \qquad \varepsilon_{\mu \nu} n^{\mu \nu} = 0, \qquad \varepsilon_{\mu \nu} \varepsilon^{\mu \nu} = -2.
\end{align}
In particular, both $n$ and $\varepsilon$ are degenerate as forms ($n \wedge n=\varepsilon \wedge \varepsilon=0)$.  We may now define projectors parallel and perpendicular to the foliation,
\begin{align}
    h_{\mu \nu} = - \varepsilon_{\mu \rho} \varepsilon_{\nu}{}^{\rho}, \qquad h^\perp_{\mu \nu} = n_{\mu \rho} n_{\nu}{}^{\rho}. \label{hdef} 
\end{align}
The projector $h$ agrees with the induced metric on the worldsheet.\footnote{In Ref.~\cite{Grozdanov:2016tdf}, our $\varepsilon$ was denoted $u$, our $h$ was denoted $\Omega$, and our $h_\perp$ was denoted $\Pi$.} The spacetime metric and volume element are reconstructed as
\begin{align}
g_{\mu \nu} = h_{\mu \nu} + h^\perp_{\mu \nu}, \qquad
\epsilon_{\mu \nu \rho\sigma} = 6 \varepsilon_{[\mu \nu} n_{\rho \sigma]}.
\end{align}
The latter equation is simply $\epsilon = \varepsilon \wedge n$.  These orientation choices are consistent with $\epsilon = dt \wedge dx \wedge dy \wedge dz$, $\varepsilon=dt\wedge dz$, and $n=dx\wedge dy$.  

The binormal is {\it almost} invariant under string relabelings -- it transforms as as $n \to \pm n$, where $\pm$ is the sign of the Jacobian determinant \eqref{jac}.  The volume element $\varepsilon$ is likewise invariant only up to sign, so a fully invariant quantity must involve an even number of total appearances of $\varepsilon$ and $n$.  This restriction corresponds to the lack of a preferred orientation of the worldsheets.  The projectors $h$ and $h_\perp$ are completely invariant and may appear in any number.

The worldsheet photon $a_\mu$ may only appear in the following combination:
\begin{align}\label{thingy}
    \tilde{f}_{\mu \nu} = h^\rho{}_\mu h^\sigma{}_\nu \le(\p_{\rho}a_{\sig} - \p_{\sig} a_{\rho} - b_{\rho\sig}\ri)
\end{align}
  The particular combination $da - b$ is manifestly invariant under the 1-form gauge transformation \eqref{ba}. The projectors guarantee invariance under the string-dependent shift \eqref{chem-shift}, as from \eqref{hdef} and \eqref{epdef} we see that $h^\rho{}_\mu \p_{\rho} \Phi_I = 0$. We refer to $\tilde{f}$ as the worldsheet field strength.  


We note that $b_{\mu \nu}$ may appear in the combination \eqref{thingy} as well as in terms of the three-form $db$, which is separately invariant under \eqref{ba}. 

\subsection{The field equations are conservation laws}

We now consider the most general action $S[\Phi_I, a; g,b]$ respecting these symmetries.  The field equations are obtained by varying with respect to the dynamical variables $\Phi_I$ and $a$.  However, the symmetries have been chosen so that the result is  \textit{equivalent} to demanding conservation of the currents obtained by varying with respect to the non-dynamical sources.  That is, the field equations of this theory are simply
\be
\nabla_\mu T^{\mu \nu} = \ha (db)^{\nu}{}_{\rho \sigma} J^{\rho \sigma}, \qquad \nabla_\mu J^{\mu \nu}=0, \label{field-as-cons} 
\ee
where
\be
T_{\mu\nu} \equiv \frac{2}{\sqrt{-g}} \frac{\delta S}{\delta g_{\mu\nu}} \qquad J^{\mu\nu} \equiv \frac{2}{\sqrt{-g}} \frac{\delta S}{\delta b_{\mu\nu}}. \label{TJdef2}
\ee
When the external charge $db$ is vanishing, we obtain precisely the conservation laws \eqref{currents}.
 
To prove these claims, let $\delta_\Lambda$ represent the variation with respect to an infinitesimal magnetic photon shift \eqref{ba}.  Then the action varies as
\be
\delta_{\Lam} S[\Phi_I, a; g, b] = \int d^4x \le(2 \frac{\delta S}{\delta b_{\mu\nu}} \p_{[\mu} \Lam_{\nu]} + \frac{\delta S}{\delta a_{\mu}} \Lam_{\mu}\ri).
\ee
By construction, this variation vanishes for {\it any} field configuration; thus after an integration by parts, and using the definition of the flux tensor \eqref{TJdef2} we conclude that 
\be
\frac{1}{\sqrt{-g}} \frac{\delta S}{\delta a_{\nu}} =  \nabla_{\mu} J^{\mu\nu}. \label{avar} 
\ee
Thus the $a_\mu$ field equation is equivalent to the conservation of magnetic flux.   Relations of this sort are familiar from the physics of Goldstone modes, and generically appear whenever a symmetry is non-linearly realized, as in the transformation of $a$ in \eqref{ba}.  

Varying instead by an infinitesimal diffeomorphism, we now find (see details in Appendix \ref{app:diffvar}) 
\begin{align}
\frac{1}{\sqrt{-g}}\frac{\delta S}{\delta \Phi_I} \nabla_\sigma \Phi^I &  = \nabla_{\mu} T^{\mu}_{\phantom{\mu}\sig}   - \ha J^{\mu\nu} (db)_{\sig\mu\nu} \nonumber \\
& +  \nabla_{\mu}  J^{\mu\nu}\le( b_{\sig\nu} - \nabla_{\sig} a_{\nu} + \nabla_{\nu} a_{\sig} \ri). \label{diffvarbulk} 
\end{align}
This shows that imposing both field equations implies the conservation laws \eqref{field-as-cons}, but we must establish the converse for full equivalence.  Projecting parallel to the foliation gives
\begin{align}
\bigg(\nabla_{\mu} T^{\mu}_{\phantom{\mu}\sig} &  - \ha J^{\mu\nu} (db)_{\sig\mu\nu} \nonumber \\
+  \nabla_{\mu} & J^{\mu\nu}\le( b_{\sig\nu} - \nabla_{\sig} a_{\nu} + \nabla_{\nu} a_{\sig} \ri)\bigg) h^{\sig}{}_{\rho} = 0 \label{Tpar}
\end{align}
while projecting perpendicular gives 
\begin{align}
\frac{1}{\sqrt{-g}}\frac{\delta S}{\delta \Phi_I} \nabla_\sigma \Phi^I &  = \bigg( \nabla_{\mu} T^{\mu}_{\phantom{\mu}\sig}   - \ha J^{\mu\nu} (db)_{\sig\mu\nu} \\
& +  \nabla_{\mu}  J^{\mu\nu}\le( b_{\sig\nu} - \nabla_{\sig} a_{\nu} + \nabla_{\nu} a_{\sig} \ri) \bigg) h_\perp{}^\sigma{}_\rho. \nonumber
\end{align}
Now the point is that if the foliation is regular ($d\Phi_1 \wedge d\Phi_2 \neq 0$), then $d\Phi_1$ and $d\Phi_2$ are a good basis for the perpendicular (co)tangent space.  This means that imposing the field equations \eqref{field-as-cons}, which causes the right-hand-side to vanish, forces each $\delta S /\delta \Phi_I$ to vanish individually.  This proves the equivalence.

This discussion reveals that the field equations provide only the perpendicular components of stress-energy conservation; the parallel ones come for free as an identity \eqref{Tpar}.  This is true even in ordinary FFE, where it has been unappreciated (at least by us). Formally,  it may be understood as a consequence of the fact that diffeomorphisms in the worldsheet directions do not act on $\Phi_I$, i.e. if $\xi^{\mu} = h^{\mu}_{\phantom{\mu}\nu} \xi^{\nu}$ then $\sL_{\xi} \Phi_I = \p_{\mu} \Phi_I h^{\mu}_{\phantom{\mu}\nu} \xi^{\nu} = 0$. This is a somewhat unfamiliar statement for local field theories---usually all dynamical degrees of freedom transform under diffeomorphisms, and clearly it is the fact that we are correlating the diffeomorphism with the state of the system through $h^{\mu}_{\phantom{\mu}\nu}$ that makes this possible. 

The upshot of this discussion is that we need never vary the action with respect to $\Phi$, $a$; we may obtain a full description of the dynamics purely from the conservation of the conserved currents, as usual in hydrodynamics. 

\subsection{Leading order: generalized FFE} \label{sec:ideal}
We now write down the most general action to leading order in derivatives.  Since we want the foliation to be the fundamental object, it is $d\Phi$ that is zeroth order in derivatives and not $\Phi$ itself.  The foliation invariants (binormal, volume element, projectors) are similarly zeroth order.
We would also like to allow an equilibrium configuration with a nonzero magnetic flux, i.e. $J \sim O(\p^0)$. This requires that we similarly take $\tilde{f}$ and the source $b_{\mu\nu}$ to be zeroth order in derivatives.\footnote{Note that for a given quantity the order in derivatives (which depends on the choice of dynamics) does not necessarily coincide with the engineering mass dimension (e.g. $d\Phi$ has mass dimension $1$ but is taken to be zeroth order in derivatives).} 

We will refer to the zeroth order system as {\it ideal}, adapting the terminology from hydrodynamics.  At ideal order the only scalar we can make is $\tilde{f}_{\mu \nu} \tilde{f}^{\mu \nu}$ (and functions thereof).  We denote this scalar as
\begin{align}\label{thing1}
\mu^2 = - \ha \tilde{f}_{\mu \nu} \tilde{f}^{\mu \nu}.
\end{align}
We can fix the sign of $\mu$ if we adopted a preferred orientation (choice of $\varepsilon$) on the worldsheet.  As a two-form in a two-dimensional space (the string worldsheet), $\tilde{f}_{\mu \nu}$ is proportional to the volume element, so we have
\begin{align}\label{thing2}
    \tilde{f}_{\mu \nu} =  \mu \varepsilon_{\mu \nu}.
\end{align}
Noting that $\varepsilon_{\mu \nu}$ takes care of the projection into the sheet, we can also write the explicit formula
\begin{align}
    \mu = \ha\vep^{\mu\nu}\le( b_{\mu\nu} - \p_{\mu}a_{\nu} + \p_{\nu} a_{\mu}\ri). \label{mudef} 
\end{align}
We will generally find it convenient to work with $\mu$ rather than $\mu^2$, but it should be borne in mind that all terms in the action must be invariant under the reversal $\varepsilon \to - \varepsilon$.  In particular, scalars made from $\mu$ alone must be even functions of $\mu$.  Allowing odd functions to appear would constitute some kind of chiral theory that we do not explore in this paper.

This discussion shows that, at ideal order, the action can be an arbitrary even function $p(\mu)$ of the scalar $\mu$:
\be\label{S0}
S_0[\Phi, a;g,b] = \int d^4x \sqrt{-g} \ \! p(\mu).
\ee
We will shortly show that a particular choice of $p(\mu)$ results in dynamics that is exactly equivalent to usual force-free electrodynamics.  However, let us first keep the function $p(\mu)$ arbitrary and construct the magnetic flux $J^{\mu\nu}$ and stress tensor $T^{\mu\nu}$ by varying the action with respect to $b_{\mu\nu}$ and $g_{\mu\nu}$ respectively. A helpful intermediate result is $\delta_{g} \mu = \ha \mu h_{\al\beta} \delta g^{\al\beta}$. After a short computation we find:
\be
J^{\mu\nu} = \rho \vep^{\mu\nu} \qquad T^{\mu\nu} = p g^{\mu\nu} - \mu \rho h^{\mu\nu}, \label{tj0}
\ee
where we have defined the scalar $\rho$
\be
\rho(\mu) \equiv \frac{d p}{d \mu}.\label{rhodef}
\ee
From \eqref{tj0}, we see that $\rho$ measures the magnitude of magnetic flux.  Eq.~\eqref{rhodef} is the zero-temperature limit of the first law of thermodynamics \cite{Grozdanov:2016tdf}, with $\mu$ the potential conjugate to the flux $\rho$. Note that at this (ideal) order we have
\be \label{thing3}
J^{\al\beta} = \frac{1}{\mu} \frac{dp}{d\mu} \tilde{f}^{\al\beta},
\ee
i.e. the current is proportional to the worldsheet field strength. 

While the action principle involves $(\Phi_1,\Phi_2,a_\mu)$, the resulting field equations are just the conservation \eqref{field-as-cons} of the currents \eqref{tj0}.  We are free to regard some alternative set (such as $\mu$ and $\varepsilon^{\mu \nu}$) as the dynamical variables when solving or analysing the equations.


Though this action formulation is new, this theory has been constructed before (at least in the absence of the external current $db$).  To our knowledge, it first appeared in \cite{Freytsis:2015qda} as a generalization of force-free electrodynamics, obtained by applying the usual force-free arguments to a non-linear theory of electromagnetism.
(We demonstrate the equivalence in App.~\ref{app:FFNE}.)  In \cite{Grozdanov:2016tdf}, the same equations of motion were constructed using higher-form symmetries as a ``zero-T'' version of magnetohydrodynamics that respected Lorentz boosts along magnetic field lines. Finally, as this paper was nearing completion, \cite{Glorioso:2018kcp} appeared, in which a similar zero-T limit is realized as a symmetry enhancement of a different action principle for MHD. 

\subsubsection{FFE as the scale-free ideal theory}

To obtain ``normal'' FFE, consider expanding $p(\mu)$ in powers of $\mu$. As we require $p$ to be an even function, we find:
\be\label{p-expansion}
p(\mu) = \frac{1}{2} \mu^2 + \frac{1}{M^2} \mu^4 + \cdots 
\ee
where $M$ is a quantity with dimensions of mass.  (Since we treat the action classically, the overall scale doesn't matter and with foresight we have set the leading coefficient to $1/2$.)  If we imagine a system where the dimensionful quantity $\mu$ is much smaller than any other quantity in the problem, then we are justified in neglecting all terms other than the first.  This is the only term with no dimensionful parameters, and it turns out to correspond to FFE,
\be
p(\mu) = \ha \mu^2 \qquad \textrm{(FFE)}.\label{pFFE}
\ee
To see the equivalence we use the map $J=\star F$ \eqref{Fdef},
\begin{align}
    \rho \varepsilon^{\mu \nu} = \tfrac{1}{2} \epsilon^{\mu \nu \rho \sigma} F_{\rho \sigma} .\label{J=starF}
\end{align}
This makes the foliation agree with the worldsheets of the magnetic field lines.  The conservation of $J^{\mu \nu}$ is now equivalent to the no-monopoles equation $\nabla_{[\mu }F_{\rho \sigma]}=0$ \eqref{FFE}.  The degeneracy \eqref{degenerate} and magnetically dominated \eqref{magdom} constraints follow from  $\varepsilon \wedge \varepsilon=0$ and $\varepsilon^2=-2$, respectively.  

The remaining equation of FFE is $F_{\sig\nu} \nabla_{\mu} F^{\mu\nu} = 0$, or equivalently the conservation of the Maxwell stress-energy tensor.  Using Eqs.~\eqref{J=starF}, \eqref{pFFE},  \eqref{rhodef}, and \eqref{thing2} we see that\footnote{Notice that the action \eqref{S0} for this theory is just the standard Maxwell action $(-1/4)F^2=(1/4)\tilde{f}^2$.  Thus FFE follows from the Maxwell action varied with respect to foliation degrees of freedom and the worldsheet magnetic photon.}
\begin{align}\label{stuff}
    \rho = \mu = B_0, \qquad p=\tfrac{1}{2} B_0^2, \qquad \tilde{f} = \star F,
\end{align}
where $B_0=\pm \sqrt{F_{\mu \nu} F^{\mu \nu}/2}$ is the magnetic field strength in a frame with no electric field.  In light of the stress-tensor \eqref{tj0}, we see that $p=B_0^2/2$ is just the standard notion of \textit{magnetic pressure}, while $\mu \rho = B_0^2$ is the \textit{magnetic tension} along field lines.  Indeed, Eq.~\eqref{tj0} with the substitutions \eqref{stuff} is precisely the Maxwell stress-tensor of a degenerate Maxwell field (e.g. Eq.~15 of \cite{Gralla:2014yja}), establishing the full equivalence with FFE.


We conclude that conventional force-free electrodynamics is the unique scale-free theory describing the infrared dynamics of cold string fluids.  This symmetry-based derivation may help explain the ubiquity of force-free field configurations in a variety of physical contexts.


\subsection{Higher derivative terms} \label{sec:higherderiv} 

We now turn to discussion of higher derivative terms.  We first note that there are no allowed terms at odd order in the derivative expansion.  This follows from the fact that all independent zeroth-order objects ($\tilde{f}_{\mu \nu}$, $\varepsilon_{\mu \nu}$, $b_{\mu \nu}$, $g_{\mu \nu}$) have two spacetime indices.

We will be mainly interested in the case of no external current, and in the remainder of this section we always $db$ to zero after variation.  It is then helpful to separately consider three types of terms in the Lagrangian:
\begin{enumerate}
    \item Terms not involving $db$. \vspace{-.2cm}
    \item Terms linear in $db$.\vspace{-.2cm}
    \item Terms non-linear in $db$.
\end{enumerate}

The type 1 terms may involve $\tilde{f}_{\mu \nu}$ and the foliation invariants.  Alternatively, we may build all such terms from $\varepsilon_{\mu \nu}$ and $\mu$.  Varying with respect to $b_{\mu \nu}$ gives
\begin{align}
    \delta\varepsilon_{\mu \nu} = 0, \qquad \delta \mu = \frac{1}{2} \varepsilon^{\mu \nu} \delta b_{\mu \nu}.
\end{align}
This shows that for terms of type 1, the variation is always proportional to $\varepsilon$,
\begin{align}
\textrm{type 1: } \qquad J^{\mu \nu} \propto \varepsilon^{\mu \nu}.
\end{align}
Geometrically, this means that the associated flux lies in the foliation, and algebraically, it sets $J \wedge J=0$ (since $\varepsilon \wedge \varepsilon=0$).  Using the map $\star F=J$ \eqref{Fdef}, we see that type 1 terms preserve $\mathbf{E} \cdot \mathbf{B} = 0$ and hence do not accelerate particles.

The type 2 terms take the special form
\begin{align}\label{Lambda}
    \frac{1}{6}\Lambda^{[\mu \nu \rho]} (db)_{\mu \nu \rho},
\end{align}
where $\Lambda$ is any three-form built from the invariants (e.g. $\varepsilon_{\mu \nu}$ and $\mu$).  This term does not contribute to the stress-tensor as $db$ is metric-independent and set to zero after variation.  However, it does contributes to the flux current as
\begin{align}
\textrm{type 2: } \qquad T^{\mu \nu} = 0, \qquad J^{\mu \nu} = - \nabla_{\sig}\Lambda^{[\sigma \mu \nu]}. \label{type2}
\end{align}
This current is identically conserved---in fact it is the most general identically conserved two-form that can be built from $\tilde{f}_{\mu\nu}$ and the foliation invariants.  It thus alters the relationship between the magnetic flux $J^{\mu\nu}$ and the dynamical degrees of freedom without altering the dynamics itself; in this regard it is perhaps similar to the Hall conductivity for ordinary fluids, which is also an identically divergenceless contribution to a usual $U(1)$ current.  In particular, noting the map $\star F=J$ \eqref{Fdef}, a type 2 term will generically give rise to non-zero $\mathbf{E} \cdot \mathbf{B}$.

Finally, the type 3 terms give no contribution at all since we set $db=0$ after the variation,
\begin{align}
    \textrm{type 3: } \qquad T^{\mu\nu} = 0, \qquad J^{\mu \nu}=0.
\end{align}


\subsection{A second-order term introducing non-zero $\mathbf{E}\cdot \mathbf{B}$}\label{sec:specialterm}

Although ultimately one may hope to scour the full space of corrections for signs of instabilities or other observationally relevant phonemona, in this paper we confine ourselves to the lowest-hanging fruit: the introduction of non-zero $\mathbf{E}\cdot \mathbf{B}$.  Following the discussion in subsection \ref{sec:higherderiv}, this occurs in our theory only for terms of a ``topological'' character---they affect only the coupling to external particles, but not the dynamical equations.  A simple such term at second order in derivatives is
\be
S_{R} = \frac{1}{2}\int d^4x \sqrt{-g} R(\mu)  \nabla^{\al} \varepsilon^{\beta\ga} (db)_{\al\beta\ga}\label{Rcorr}
\ee
Here $R(\mu)$ is an \textit{odd} function of $\mu$ so that the term is invariant under the change of worldsheet orientation $\ep \to -\ep$ (implying $\mu \to -\mu$).  This is just the choice $\Lambda =R d\varepsilon$ in \eqref{Lambda}.  Setting $db=0$ after variation we find
\be
J^{\mu \nu}_{R} = - 3\nabla_{\sig}\le(R(\mu) \nabla^{[\sig} \varepsilon^{\mu \nu]}\ri). \label{EBcorr} 
\ee
If we expand $R(\mu)$ in terms of some mass scale $M$ as in Eq.~\eqref{p-expansion},
\be
R(\mu) = \frac{\mu}{M^2} + \frac{c_3}{M^4} \mu^3 + \cdots ,
\ee
then keeping the leading term gives the simple correction
\begin{align}\label{JR1}
J^{\mu \nu}_{R_1} = -\frac{3}{M^2} \nabla_{\sig}\le(\mu \nabla^{[\sig} \varepsilon^{\mu \nu]}\ri). 
\end{align}
As ordinary FFE corresponds to the leading term in the expansion \eqref{p-expansion} for $p$, this is the natural correction (of the form \eqref{Rcorr}).  
In Section \ref{sec:observations} we will explore the observational consequences of this term.



\subsection{Comparison with other approaches}

Finally, we comment on how our approach to higher derivative corrections contrasts with that taken previously by \cite{Grozdanov:2016tdf}. In that work some possible second-order corrections to $T^{\mu\nu}$ and $J^{\mu\nu}$, linearized about a homogenous equilibrium configuration, were constructed. However an obstruction to their approach arose from the fact that they worked not with an action but rather directly with $\ep^{\mu\nu}$ and $\mu$. The resulting equations of motion are generically overdetermined, and consistency of the dynamical equations required that the currents obeyed a particular constraint that projected out two of the degrees of freedom. This constraint was only precisely formulated for the ideal order system -- in the language of this paper, it is \eqref{Tpar}, which when restricted to the ideal order system \eqref{tj0} becomes 
\be
\le(\nabla_{\mu} T^{\mu}_{\phantom{\mu}\sig} + \mu \nabla_{\mu} J^{\mu\nu} \ep_{\nu\sig}\ri)h_{\sig\al} = 0 ,
\ee
where details are given in Appendix \ref{app:diffvar}. In \cite{Grozdanov:2016tdf}, it was technically difficult to generalize this constraint to higher orders in derivatives, and thus it remained unclear whether the higher derivative corrections written down there were actually consistent beyond the linearized level where the consistency of the resulting set of equations could be verified directly.  

In our formalism, the constraint arises from the off-shell realization of diffeomorphisms along the worldsheet, and its generalization to all orders in derivatives is given in \eqref{Tpar}.  It is satisfied automatically and plays no role in the analysis.  It would be very interesting to systematically classify all terms that can arise from our action approach and compare them with the set of terms written in \cite{Grozdanov:2016tdf}. 

Finally, we note some possible limitations of our action-based approach. It is well-known that any system described by a conventional action cannot describe dissipation, and thus the approach of (e.g.) \cite{Dubovsky:2011sj} to conventional finite-$T$ hydrodynamics does not allow for the appearance of any dissipative transport coefficients, which of course are allowed to appear in formulations of hydrodynamics based purely on the equations of motion. Indeed, a great deal of recent work in the field (see e.g. \cite{Endlich:2012vt, Grozdanov:2013dba,Kovtun:2014hpa,Harder:2015nxa,Crossley:2015evo,Haehl:2015foa,Haehl:2015uoc,Torrieri:2016dko}) has resulted in the construction of much more sophisticated action principles. The state of the art now allows for dissipation and a systematic treatment of fluctuations, but requires a doubling of fundamental fields, with the partner fields basically now living on the two branches of the Schwinger-Keldysh contour that is used in finite-temperature real-time quantum field theory. 

Returning to FFE, the Lorentz-invariance of our system seems to forbid the existence of dissipation in the traditional sense; nevertheless, one may wonder whether our single action is  missing some particular class of higher-derivative terms that could generically be present in an approach based only on the equations of motion. Indeed, there is evidence that the simple action formulation may even miss some {\it non}-dissipative terms in finite-$T$ hydrodynamics \cite{Bhattacharya:2012zx}. One might hope to clarify such issues (as well as to understand how to incorporate fluctuations) by applying the Schwinger-Keldysh approach described above to our framework. 

Finally, we note that a different approach to producing a non-zero $\mathbf{E} \cdot \mathbf{B}$ in astrophysical settings is to introduce a non-zero resistivity into the system.  This requires the selection of a rest frame, which can be done by introducing a non-zero temperature (see e.g. \cite{Grozdanov:2016tdf,Hernandez:2017mch}) or, for spacelike currents, by demanding vanishing charge density and parallel electric and magnetic fields \cite{gruzinov2008}.  
Resistive pulsar magnetospheres were studied in Refs.~\cite{li-spitkovsky-tchekhovskoy2011,kalapotharakos-kazanas-harding2012}.  Our construction does not appear to be simply related to these ideas and our expressions (though covariant) do not agree with those of \cite{gruzinov2008}.  We obtain nonzero $\mathbf{E} \cdot \mathbf{B}$ even though our dynamics are completely dissipationless.  We believe the physics underlying Eq.~\eqref{JR1} has little to do with resistivity as conventionally defined. 

\section{Solutions}\label{sec:solutions}

We now consider some simple solutions to clarify the physics of our description.

\subsection{Homogeneous Field}

We first discuss the simplest example solution to the ideal theory (generalized FFE).  To find a solution we first select the external sources $g_{\mu \nu}$ and $b_{\mu \nu}$.  We wish to consider flat spacetime with no external charge-current, so we take 
\begin{align}
    ds^2 = -dt^2 + dx^2 + dy^2 + dz^2, \qquad b_{\mu \nu}=0.
\end{align}
We are of course free to apply diffeomorphisms and one-form shifts ($b \to b+ d\Lambda$) without altering the external environment.

A simple timelike foliation are the strings at fixed $x$ and $y$.  We may represent this foliation by $\Phi_1=x$ and $\Phi_2=y$.  The remaining variable in the action formulation is $a_\mu$, which will solve the field equations if we also take $a=zdt$.  That is, a simple solution for the dynamical variables is 
\be
\Phi_1 = x, \qquad \Phi_2 = y, \qquad a= -\mu_0 zdt,
\ee
where $\mu_0$ is a constant. Of course, there is tremendous gauge freedom and it is more helpful to consider invariants.  The binormal and volume elements are
\begin{align}
    n = d x \wedge d y, \qquad \varepsilon = dt \wedge dz, 
\end{align}
and the gauge-invariant ``worldsheet field strength'' is
\begin{align}
    \tilde{f} =\mu_0 dt \wedge dz. \
\end{align}
We note that by a magnetic photon shift \eqref{ba}, $a$ can be set to $0$ at the cost of turning on a source $b$, but of course $\tilde{f}$ remains unchanged. From \eqref{mudef} we have $\mu = \mu_0$. 

From \eqref{tj0} we find:
\be
J = \frac{dp}{d\mu}\bigg|_{\mu = \mu_0} dt \wedge dz,
\ee
corresponding to a magnetic field pointing in the $z$ direction. Note that the precise value of the field depends on the equation of state $p(\mu)$.  For the conventional FFE with equation of state \eqref{pFFE}, we this is a constant magnetic field $J = \mu_0 dt \wedge dz$.

\subsection{Michel Monopole}

We now consider a more complicated solution, the Michel monopole \cite{michel1973mon} of (ordinary) FFE, i.e. with $p(\mu) = \ha \mu^2$.  This solution represents the exterior magnetosphere of a rotating, conducting sphere (in flat spacetime) that has been magnetized such that the radial component of magnetic field is uniform over the sphere.  The field strength is given in spherical coordinates by (e.g. \cite{Gralla:2014yja})
\begin{align}
    F = q \sin \theta d\theta \wedge (d\phi - \Omega d(t-r)).
\end{align}
Here $q$ is the magnetic monopole charge of the solution; however, in applications we would multiply the solution by $\textrm{sign}(\cos \theta)$ to ``split'' the monopole.  This eliminates the actual monopole charge while introducing a current sheet along the equator that mimics the phenomenology of a more realistic pulsar (see e.g. \cite{Gralla:2014yja} for discussion). 

To express in our language we compute the two-form current $J^{\mu \nu}$ \eqref{Fdef},
\begin{align}\label{JMichel}
   J = \star F = \frac{q}{r^2} d(t-r) \wedge (dr - r^2 \Om \sin^2 \th d\phi).
\end{align}
As $J=\mu \varepsilon$ in ordinary FFE (Eqs.~\eqref{J=starF} and \eqref{stuff}), we thus have
\be\label{muepsMichel}
\varepsilon = d(t-r) \wedge (dr - r^2 \Om \sin^2 \th d\phi) \qquad \mu = \frac{q}{r^2}.
\ee
An allowed choice of the action degrees of freedom is 
\begin{align}
    \Phi_1=\theta, \qquad \Phi_2=\phi-\Omega(t-r), \qquad a= \frac{q}{r} dt
\end{align}
with the metric flat and in spherical coordinates and the source $b$ vanishing.  These choices make clear the interpretation of a \textit{rotating monopole}---we have the vector potential $a_\mu$ of a monopole attached to a rotating foliation $\Phi_1$ and $\Phi_2$.  This is an ordinary charge for the \textit{magnetic} photon, so it manifests physically as a magnetic monopole.  

\section{Observational Consequences} \label{sec:observations} 
We now give a brief discussion of potential observational consequences, focusing on the simple second-order correction \eqref{Rcorr}.  The qualitative effect of this term, relative to the ideal background theory, is the introduction of non-zero $\mathbf{E}\cdot \mathbf{B}$.  This gives rise to particle acceleration along magnetic field lines, potentially producing observed particle winds.  We consider the Michel solution for definiteness, but the main conclusions hold for force-free pulsar magnetospheres more generally.


We consider ordinary FFE (the ideal theory \eqref{S0} with $p=\tfrac{1}{2}\mu^2$) corrected by \eqref{Rcorr} with the leading term $R(\mu)=\mu/M^2$.  This theory has a single scale, the unknown mass $M$.  (As we set $\hbar=c=1$, this can equivalently be thought of as a length or a timescale.)  The correction is ``topological'' in that it affects only the \textit{map} between the degrees of freedom and and the magnetic field, and not the equations of motion themselves.  Thus Michel solution continues to solve the corrected equations of motion, but the flux tensor (dual electromagnetic field) undergoes a shift,
\begin{align}\label{Jcorr}
    J = J_{\rm Michel} + J_{R_1},
\end{align}
where $J_{\rm Michel}$ is given in \eqref{JMichel}
and $J_{R_1}$ is computed from Eq.~\eqref{JR1} using the the Michel values \eqref{muepsMichel} of $\mu$ and $\varepsilon$.  As $J$ is just $\star F$, the invariants may be constructed as,
\begin{align}
    \mathbf{E} \cdot \mathbf{B} & = \tfrac{1}{8} \epsilon_{\mu \nu \rho \sigma} J^{\mu \nu} J^{\rho \sigma} \\
    \mathbf{B}^2-\mathbf{E}^2 & = -\tfrac{1}{2} J_{\mu \nu} J^{\mu \nu}.
\end{align}
We are mainly interested in the effective electric field along field lines,\footnote{This definition is suitable when $\mathbf{E} \cdot \mathbf{B} \ll \mathbf{B}^2-\mathbf{E}^2$, as occurs in this perturbative calculation.} 
\begin{align}
    E_0 \equiv \frac{\mathbf{E} \cdot \mathbf{B}}{\sqrt{\mathbf{B}^2-\mathbf{E}^2}}.
\end{align}
Using \eqref{Jcorr}, \eqref{JMichel} and \eqref{JR1}, we find to leading order in $1/M$ that
\begin{align}
    E_0 = -\frac{4 q}{M^2 r^3} \Omega \cos\th .
\end{align}
We will now work with more astrophysically convenient quantities (recall that $\hbar=1$)
\begin{align}
    q & = B_* R_*^2 \\
    M & = 1/L.
\end{align}
Here $B_*$ is the magnetic field at the stellar radius $R_*$, and and $L$ is the microscopic lengthscale of the theory.  We may then write
\begin{align}
    E_0 = -4 B_\star (\Omega R_*)\left(\frac{R_*}{r}\right)^3 \left(\frac{L}{R_\star} \right)^2 \cos \theta \label{E0}
\end{align}
As the field lines are straight, we may integrate radially to determine the voltage between the stellar surface and infinity,
\begin{align}\label{V}
    V =  2 V_0 \cos \theta \left( \frac{L}{R_*} \right)^2, \qquad V_0= B_* R_*^2 \Omega,
\end{align}
where $V_0$ is the typical ``unipolar inductor'' voltage generated by the rotation of the star through the magnetic flux $B_* R_*^2$.  This is a famously large voltage for pulsars,
\begin{align}
    V_0 \approx (6 \times 10^{16} \textrm{ Volts}) \ \! \frac{B_{12}}{P_1},
\end{align}
where $B_{12}$ is the stellar magnetic field in units of $10^{12}$ Gauss, while $P_1$ is the pulsar period in seconds.  (We use a stellar radius of ten kilometers.)  The actual voltage is down by two powers of the dimensionless ratio $L/R_*$.  The typical Lorentz factor of an electron is then (assuming $\gamma \gg 1$ and dropping factors)
\begin{align}\label{gamma}
    \gamma \sim 10^{11} \epsilon_r \frac{B_{12}}{P_1} \left( \frac{L}{R_*} \right)^2,
\end{align}
where we have used the mass of the electron $m_e \approx 0.5\;\mbox{MeV}$ and where $\epsilon_r<1$ is some efficiency factor reflecting radiative and other losses during acceleration. 

Although Eq.~\eqref{gamma} was derived in the case of purely radial field lines (the Michel monopole model), we expect it to hold for pulsar magnetospheres more generally.  From \eqref{JR1} we see that the correction to the magnetic field scales as $\mathcal{B} L^2/\mathcal{R}^2$, where $\mathcal{B}$ and $\mathcal{R}$ are typical scales of magnetic field strength and variation, respectively.  Near the star these can be identified with the stellar magnetic field and radius, up to model-dependent factors.  Then the electric field correction $E_0$ will take the form of \eqref{E0} with $(R_*/r)^3$ replaced by some more complicated function of $r/R_*$ characterizing the fall-off of the field.  This will affect the voltage \eqref{V} only by numerical factors.  Indeed, $V \propto V_0$ can be expected on purely physical grounds, as $V_0$ is the typical voltage of a unipolar inductor.  We conclude that the estimate \eqref{gamma} is largely insensitive to the details of the magnetic field configuration.

We can use Eq.~\eqref{gamma} to estimate the scale $L$ required to match some given observed Lorentz factor $\gamma$.  Using a ten-kilometer stellar radius and solving for $L$ yields
\begin{align}
    L \sim (10^{-5} \textrm{km})\sqrt{\frac{\gamma}{\epsilon_r}} \sqrt{\frac{P_1 }{B_{12}}}.
\end{align}
The right-most factor $\sqrt{P_1/B_{12}}$ ranges from $\sim .05$ to a few, with the majority of pulsars around 1/2.  Suppose that we wish to reproduce a Lorentz factor $\gamma \sim 10^3$ of a robust pulsar wind nebula (e.g. \cite{gaensler-slane-PWN-review2006}), and assume an efficiency of $\epsilon_r \sim .1$.  Then the lengthscale $L$ comes out to about a meter.  As a frequency this is 300 MHz, which happens to be right where the pulsar \textit{radio} spectrum often peaks (e.g. \cite{pulsar-spectra}).  This suggests the tantalizing possibility that a single EFT, with a scale $L\sim$meters, could explain the disparate phenomena of radio emission and pulsar wind.  Needless to say, however, significant further work is required before the EFT could be considered a viable model of either effect.

We conclude by discussing how the coefficients in an effective theory (e.g. the function $R(\mu)$ in \eqref{Rcorr}, or more specifically the precise value of the scale $M=1/L$) can in principle be computed from a microscopic description. In theories of conventional hydrodynamics, this is done through {\it Kubo formulas} that relate hydrodynamic transport coefficients to two-point correlation functions of the conserved currents in thermal equilibrium (see e.g. \cite{Kovtun:2012rj} for a modern review).  It is possible to derive similar Kubo formulas in our framework, which would relate the coefficient $M$ to 2-point functions of the magnetic flux and stress tensor operators in magnetic equilibrium. These could then in principle be computed from microscopic quantum field theory, or (more realistically) from numerical simulation of homogenous strong-field plasma.  In this way the EFT could bridge the gap between a microscopic description and macroscopic observations.  




\section*{Acknowledgements}
It is a pleasure to acknowledge helpful discussions and correspondence with S. Grozdanov, D. Hofman, A. Jain, A. Nicolis and N. Poovuttikul. We are grateful to the Aspen Center for Physics (which is supported by National Science Foundation grant PHY-1607611), where the initial contact that would lead to this collaboration was made.  SG is supported in part by the NSF under award PHY-1752809 to the University of Arizona.  NI is supported in part by the STFC under consolidated grant ST/L000407/1. 

\appendix

\section{Dualizing and the magnetic photon} \label{app:magphot} 
For completeness, here we review why one might consider the transformation \eqref{ba} to be analagous to that of a magnetic photon. Similar discussion appears in the Introduction of \cite{Hofman:2017vwr} and Appendix C of \cite{Cordova:2018cvg}. 

Consider free electromagnetism written in terms of the normal electric vector potential, with no dynamical electric charges but coupled to the external source $b$ as in \eqref{qedac} i.e.
\be
S[A;b] = \int \le(-\frac{1}{2} dA  \wedge \star dA + b \wedge dA\ri). \label{freeqedac} 
\ee
Recall that this action is invariant under the 1-form shift $b \to b + d\Lam$. As only $dA$ appears in the action, we may perform electri-magnetic duality at the level of the action in the usual manner (see e.g. Appendix B of \cite{Polchinski:1998rr}). This is done by treating $F \equiv dA$ as the dynamical variable rather than $A$; however one must then introduce a Lagrange multiplier $\tilde{A}$ to enforce that $F$ is closed, i.e. we obtain:
\be
S[F,\tilde{A};b] = \int  \le(-\frac{1}{2} F \wedge \star F + b \wedge F - \tilde{A} \wedge dA \ri).
\ee
As $F$ appears only quadratically, we may eliminate it from the action by solving its equations of motion and plugging back in to find:
\be
S[\tilde{A};b] = \int  \le(-\frac{1}{2} (d\tilde{A} - b ) \wedge \star (d\tilde{A} - b)\ri) \label{magac}
\ee
$\tilde{A}$ is usually called the {\it magnetic photon}; we see that unlike in the original formulation \eqref{freeqedac} the 1-form shift now requires a compensating shift of $\tilde{A}$:
\be
\tilde{A} \to \tilde{A} + \Lam \qquad b \to b + d\Lam \label{magphot} 
\ee
This is precisely the symmetry transformation postulated for $a$ in \eqref{ba}; hence we refer to it as the ``magnetic photon shift''.  

We stress, however, that the physics of $a$ is {\it not} identical to that of the usual magnetic photon $\tilde{A}$. In particular, the extra symmetry \eqref{chem-shift} effectively confines the dynamical part of $a$ to the worldsheet. Furthermore, the route that we have taken to justify the transformation \eqref{magphot} involves a dualization that is only possible when there are no electric charges, which is certainly not the case for a plasma. Relatedly, as the action \eqref{magac} suggests, the magnetic photon $\tilde{A}$ is actually a Goldstone boson of a spontaneously broken generalized global symmetry \cite{Gaiotto:2014kfa, Hofman:2018lfz, Lake:2018dqm}, whereas this is not the case for the worldsheet photon $a$. 

Consider now the transformations of the form $\Lam = d\lam$ with $\lam$ a 0-form; in this case the transformation reduces to
\be
\tilde{A} \to \tilde{A} + d\lam
\ee
leaving the source invariant. Transformations of this sort are usually called (magnetic) $U(1)$ gauge transformations -- to be more precise, if $\lam$ has compact support then this transformation is ``pure gauge'' and does not act on the physical configuration space (i.e. in a quantum treatment it leaves all physical states invariant).  

However, if $\lam$ extends to infinity or fails to be globally well-defined (by winding, say, around a compact cycle) then this is a so-called ``large gauge'' transformation that does act on physical states. In modern language this is now the action of a generalized global symmetry \cite{Gaiotto:2014kfa,Lake:2018dqm}. These issues do not play a role in any of our analysis. 

\section{Diffeomorphism variation of action}  \label{app:diffvar} 
Here we present some details of the diffeomorphism variation of the action $S[\Phi_I, a; g, b]$. Under a diffeomorphism, all fields $\psi$ (whether external sources or dynamical fields) vary as
\be
\delta_{\xi} \psi = \sL_{\xi} \psi
\ee
where $\sL_{\xi}$ is the Lie derivative. The diffeomorphism variation of the action is then:
\begin{widetext}
\be
\delta_{\xi} S[\Phi_I, a; g, b] = \int d^4x \le(\frac{\delta S}{\delta g_{\mu\nu}} \delta_{\xi} g_{\mu\nu} + \frac{\delta S}{\delta b_{\mu\nu}} \delta_{\xi} b_{\mu\nu} + \frac{\delta S}{\delta a_{\mu}} \delta_{\xi} a_{\mu} + \frac{\delta S}{\delta \Phi_I} \delta_{\xi} \Phi^I\ri) = 0.
\ee
By construction this must vanish. Now from the definition \eqref{TJdef2} of the stress and flux tensors this may be rearranged to read:
\be
\int d^4x \le[\sqrt{-g}\le(-\nabla_{\mu} T^{\mu}_{\phantom{\mu}\sig} + \ha J^{\mu\nu} (db)_{\sig\mu\nu} -   \nabla_{\mu} J^{\mu\nu} b_{\sig\nu}\ri)\xi^{\sig} + \frac{\delta S}{\delta \Phi^I} \delta \Phi^I + \frac{\delta S}{\delta a_{\mu}} \delta_{\xi} a_{\mu} \ri].
\ee
We may further use the relation \eqref{avar} for the variation $\delta S/\delta a$ to find as an identity:
\be
\int d^4x \sqrt{-g} \le[-\nabla_{\mu} T^{\mu}_{\phantom{\mu}\sig} + \ha J^{\mu\nu} (db)_{\sig\mu\nu} +  \nabla_{\mu} J^{\mu\nu}\le(\nabla_{\sig} a_{\nu} - \nabla_{\nu} a_{\sig} -  b_{\sig\nu} \ri) + \frac{1}{\sqrt{-g}}\frac{\delta S}{\delta \Phi^I} \p_{\sig} \Phi^I\ri]\xi^{\sig} = 0. \label{fulldiff} 
\ee
\end{widetext} 
This is equation \eqref{diffvarbulk} in the text.  

We now consider a diffeomorphism parallel to the worldsheet, i.e. $\xi^{\mu} = h^{\mu}_{\phantom{\mu}\nu} \xi^{\nu}$ Putting this in \eqref{fulldiff}, we find that the last term in $\frac{\delta S}{\delta \Phi_I}$ vanishes, and we have the following off-shell identity, which is \eqref{Tpar} in the text:
\begin{align} 
\big(-\nabla_{\mu} T^{\mu}_{\phantom{\mu}\sig} &  + \ha J^{\mu\nu} (db)_{\sig\mu\nu} \nonumber \\
+  \nabla_{\mu} & J^{\mu\nu}\le( \nabla_{\sig} a_{\nu} - \nabla_{\nu} a_{\sig}  - b_{\sig\nu} +\ri)\big) h^{\sig\ga} = 0 \label{origcons}
\end{align}
This is a constraint that must hold on any stress and flux tensor obtained in our formalism. We now connect it with a constraint that played an important role in the analysis of \cite{Grozdanov:2016tdf}. To proceed, we assume (as was done in \cite{Grozdanov:2016tdf}) that we are working to ideal order. We then have
\be
J^{\mu\nu} = \rho \vep^{\mu\nu} = \frac{\rho}{ |d \Phi_1 \wedge d \Phi_2|}\le(\ep^{\mu\nu\rho\sig} \p_{\rho} \Phi_1 \p_{\sig} \Phi_2 \ri)
\ee
However, by antisymmetry the divergence of the bracketed quantity above is zero, and thus we find 
\be
\nabla_{\mu} J^{\mu\nu} = \nabla_{\mu}\le(\frac{\rho}{ |d \Phi_1 \wedge d \Phi_2|}\ri)(\ep^{\mu\nu\rho\sig} \p_{\rho} \Phi_1 \p_{\sig} \Phi_2 )
\ee
Thus the free $\nu$ index necessarily points in a world-sheet direction, and we have:
\be
\nabla_{\mu} J^{\mu\nu} = \nabla_{\mu} J^{\mu\al} h_{\al}^{\phantom{\al}\nu} 
\ee
Using \eqref{mudef} we then find that the last term in \eqref{origcons} can be written entirely in terms of $\mu$:
\be
\nabla_{\mu} J^{\mu\nu}h_{\al}^{\phantom{\sig}\nu} \le( b_{\sig\nu} - \nabla_{\sig} a_{\nu} + \nabla_{\nu} a_{\sig} \ri)h^{\sig\ga} = \mu \nabla_{\mu} J^{\mu\nu} \ep_{\sig\nu} h^{\sig\al}
\ee
Next, we turn to the term in $(db)_{\sig\mu\nu}$ in \eqref{origcons}; we note that each of the three indices $\sig, \mu, \nu$ must be different; however as $J^{\mu\nu} \sim \ep^{\mu\nu}$ they must also all point in the world-sheet directions for the term not to vanish. However there are only two such directions. Thus we find finally:
\be
\le(\nabla_{\mu} T^{\mu}_{\phantom{\mu}\sig} + \mu \nabla_{\mu} J^{\mu\nu} \ep_{\nu\sig}\ri)h_{\sig\al} = 0 
\ee
which is precisely the ideal-order constraint found by inspection in \cite{Grozdanov:2016tdf}. We now see that it is a consequence of diffeomorphisms being realized even off-shell along the field lines. There is further discussion of this point in the bulk of the text. 


\section{Equivalence of ideal order with generalized force-free electrodynamics}\label{app:FFNE}

We now explore the link between the ideal-order equations of this EFT (Sec.~\ref{sec:ideal}) and the force-free dynamics of non-linear electromagnetism \cite{Freytsis:2015qda}.  Consider an action that is an arbitrary function of the invariants $I=F_{\mu \nu} F^{\mu \nu}$ and $K=\tfrac{1}{2} \epsilon^{\mu \nu \rho \sigma} F_{\mu \nu} F_{\rho \sigma}$,
\begin{align}
    S_{\rm NLE} = \int \sqrt{-g} d^4 x \mathcal{L}(I,K).
\end{align}
This action defines a stress-tensor $T^{\mu \nu}_{\rm NLE}$ in the usual way \eqref{TJdef2}.  Demanding conservation of the stress-tensor (but \textit{not} demanding any field equations associated with $S_{\rm NLE}$) turns out to imply \cite{Freytsis:2015qda}
\begin{align}
    F_{\sig\nu} \nabla_{\mu} \le( \left. \frac{\pd \mathcal{L}}{\pd I} \right|_{K=0} F^{\mu\nu}\ri)  = 0, \label{gen1} 
\end{align}
where we have also assumed degeneracy \eqref{degenerate}.  Adjoining the no-monopoles condition,
\begin{align}
    \nabla_{[\mu} F_{\rho \sigma]} = 0, \label{gen2} 
\end{align}
constitutes the force-free dynamics of a non-linear electromagnetism.

We now show that these equations are equivalent to the ideal-order theory of Sec.~\ref{sec:ideal}.  The map is 
\begin{align}
   J^{\mu\nu} = \ha\ep^{\mu\nu\rho\sig} F_{\rho\sig}, \quad I =2 \rho^2, \quad \mathcal{L}|_{K=0} = p - \mu \rho. \label{themap} 
\end{align}
Note that $\sL$ is the Legendre transform of $p$ to a function of $\rho$ rather than $\mu$. For later use use $dp = \rho d\mu$ to show that:  
\be
\frac{\p \mathcal{L}}{\p I}\bigg|_{K = 0} = -\frac{1}{4} \frac{\mu}{\rho} \label{LImurho} 
\ee  
Using \eqref{themap}, it turns out that \eqref{gen1} and \eqref{gen2} are equivalent to the following components of our conservation equations
\be
\nabla_{\mu} J^{\mu\nu} = 0 \qquad \le(\nabla_{\mu} T^{\mu\nu}\ri)h^{\perp}_{\nu\rho} = 0
\ee
together with the definitions \eqref{tj0}.  The conservation of $J$ is trivially equivalent to \eqref{gen2}. The less trivial equation is that for $T$:
\begin{align}
\le(\nabla_{\mu} T^{\mu\nu}\ri)h^{\perp}_{\nu\rho} & = h^{\perp}_{\mu\rho} \nabla^{\mu} p - \mu\rho \vep^{\mu\sig}  h_{\nu\rho}^{\perp} \nabla_{\mu}\vep_{\sig}^{\phantom{\sig}\nu}\\
& = -\frac{3}{2} J^{\mu\nu} \nabla_{[\rho}\le(J_{\mu\nu]}\frac{\mu}{\rho}\ri) \label{lasteq}
\end{align} 
where we have used $\nabla_{\sig} p = \rho \nabla_{\sig} \mu$ as well as $\vep^{\mu\sig}\vep^{\al\beta}\nabla_{\ga} \vep_{\sig\beta} = 0$.  Using \eqref{LImurho}, the vanishing of \eqref{lasteq} is now equivalent to \eqref{gen2}. As shown in \eqref{Tpar}, the other components of the conservation of $T^{\mu\nu}$ hold as an identity and thus contain no dynamical information.

\section{Coupling an external charged particle} \label{app:externalcharge} 

Imagine that we would like to add an {\it extra} external charged particle with charge $q$ to the system.  Since our definition \eqref{Fdef} of $F$ in terms of $J$ agrees with the usual $F=dA$, we know that (neglecting plasma back-reaction) the dynamics of this particle is given by the Lorentz force law in terms of $J$:
\be
m\frac{D^2}{ds^2} X^{\sig} + \frac{q}{2} J^{\mu\nu}(X(s))  \ep_{\mu\nu\rho\sig} \dot{X}^{\rho} = 0. \label{Lorentzgeod} 
\ee
 However, it may be instructive to see this result derived from our action formulation. This is now something of a thorny point: usually in electrodynamics the action of a charged particle contains a term that is the integral of the vector potential $e \int_{C} A$ over its worldline $C$. However in our formalism the useful degrees of freedom are $(\Phi_I, a)$ and {\it not} the usual electric vector potential $A$. As we no longer have access to $A$, how can we write down the action of the particle? 

We may accomplish this following a technique used in \cite{Iqbal:2011bf} to describe the coupling of a superconducting vortex with external fields, which turns out to be a formally similar problem. We begin by recalling from \eqref{jdef} that $db$ can be viewed as an external electric charge current that is probing the system: 
\begin{align}
    j^{\sig}_{\mathrm{ext}} = - \ha \ep^{\sig\rho\mu\nu} \p_{\rho} b_{\mu\nu}.
\end{align}
Thus to add a single external charge moving on a worldline $C$, we must arrange for $j^{\sig}_{\mathrm{ext}}$ and thus $db$ to have delta function support on its worldline, i.e. we would like to construct a source field $b(x;C)$ that depends both on spacetime and a curve $C$ so that
\be
\nabla_{[\mu} b_{\nu\rho]}(x;C) = \frac{q}{2} \int_C ds \frac{dX^{\sig}}{ds} \ep_{\mu\nu\rho\sig} \delta^{(4)}\le(x - X^{\mu}(s)\ri) \label{dbeq} 
\ee
There is clearly a great deal of freedom in this choice of $b$ -- e.g. we may shift $b \to b + d\Lam$ for any 1-form $\Lam$ without affecting this equation -- but as usual this will not affect the equations of motion. For some intuition, $b$ may be considered the field strength $F$ (in ordinary Maxwell electrodynamics) of the electromagnetic field produced by a magnetic monopole living on the curve $C$. 

To obtain the dynamics of the point particle, we now consider the following combined action, which is a functional of the dynamical fields as well as of the worldline $C$
\be
S_\mathrm{pp}[\Phi_I, a, C] = m \int_C ds + S_\mathrm{FFE}[\Phi_I, a; b(x;C)]
\ee
where the first term is the usual geometric proper time along the worldline, and where $S_\mathrm{FFE}$ is the string fluid action that we have constructed. Note that the string fluid action depends on $C$ through the choice of $b(x;C)$. Denoting the particle trajectory by $X^{\mu}(s)$, we now vary $X^{\mu} + \delta X^{\mu}$ to construct the equations of motion of the particle. 

The variation of the first term gives the usual geodesic equation. Let us consider the variation of the second term:
\be
\delta_{X} S_\mathrm{FFE}[\Phi_I, a; b(x;C)] = \ha \int d^4x \sqrt{-g}  J^{\mu\nu} \delta_{X} b_{\mu\nu}(x), \label{bcvar} 
\ee
This may appear to be an extremely complicated variation, as naively varying $C$ appears to alter $b$ arbitrarily far away from the curve $C$ itself. In fact, when evaluated on an on-shell field configuration satisfying $\nabla_{\mu} J^{\mu\nu} = 0$, this variation ends up localizing on the worldline. To understand this, we note that $b(x;C)$ may be explicitly constructed by 
\be
b_{\mu\nu}(x; C) = \frac{q}{2} \int ds \dot{X}^{\rho} T_{\mu\nu\rho}(x - X(s)) 
\ee
where $T_{\mu\nu\rho}(x)$ is a ``monopole propogator'' that satisfies the following equation:
\be
\nabla_{[\al} T_{\mu\nu]\rho}(x) = \frac{1}{\sqrt{-g}}\delta^{(4)}(x) \ep_{\al\mu\nu\rho} \label{defT} 
\ee 
There is again an extremely large amount of freedom in this choice of function. One candidate is given by
\be
T_{\mu\nu\rho}(x) = \frac{3}{2} \ep_{\mu\nu\al\beta} \nabla^{[\al} G^{\beta]}_{\phantom{\beta} \rho}(x)
\ee
where $G^{\beta\rho}$ is the usual free photon propagator that satisfies:
\be
\nabla_{\al} \le(\nabla^{[\al} G^{\beta]\rho}(x)\ri) = \frac{1}{\sqrt{-g}} \delta^{(4)}(x) g^{\beta\rho}
\ee
With this choice $b_{\mu\nu}(x;C)$ is exactly the field strength produced by a magnetic monopole moving along $C$ in free electrodynamics. The precise form of $T_{\mu\nu\rho}$ will not be important for us. We may now explicitly construct the change in $b_{\mu\nu}(x;C)$ under a small variation $\delta X^{\al}$. We find
\be
\delta_{X} b_{\mu\nu}(x) = \frac{q}{2} \int ds\le(\nabla_{\rho} T_{\mu\nu\al}\le(x-X(s)\ri) - \al \leftrightarrow \rho \ri)\dot{X}^{\rho} \delta X^{\al}
\ee
Using the definition of the monopole propagator \eqref{defT} and some rearrangement of indices, this may be rewritten as 
\begin{align}
\delta_{X} b_{\mu\nu}(x) &= q \int ds \frac{dX^{\rho}}{ds} \ep_{\al\mu\nu\rho} \delta^{(4)}(x- X^{\mu}(s))  \nonumber \\
& \qquad \qquad + \nabla_{\mu} \delta c_{\nu} - \nabla_{\nu} \delta c_{\mu} \label{bvar}
\end{align}
where the first term is localized on the worldline and where $\delta c$ is an infinitesimal 1-form built out of $\delta X^{\al}$ whose form depends on the choice of $T_{\mu\nu\rho}$:
\be
\delta c_{\nu} \equiv 2q \int ds \le(T_{\al\nu\rho} - T_{\rho\nu\al}\ri) \dot{X}^{\rho} \delta X^{\al}
\ee
We now insert the expression \eqref{bvar} into  \eqref{bcvar}.  We see that the terms involving $\delta c$ above will not contribute to the equations of motion, as upon an integration by parts they result in terms proportional to $\nabla_{\mu} J^{\mu\nu}$. Only the first delta function term contributes, and as claimed the variation has localized on the worldline. Combining this with the usual geodesic equation arising from varying the proper time term, we find the full equations of motion to be \eqref{Lorentzgeod}, as claimed.

\bibliographystyle{utphys}
\bibliography{all}

\end{document}